\documentclass[lettersize,journal]{IEEEtran}
\usepackage{amsmath,amsfonts}
\usepackage{algorithmic}
\usepackage{algorithm}
\usepackage{array}
\usepackage{textcomp}
\usepackage{url}
\usepackage{verbatim}
\usepackage{graphicx}
\usepackage{subcaption}
\usepackage{cite}
\usepackage{svg}
\usepackage{colortbl}
\usepackage{graphicx}
\usepackage{multirow}
\usepackage[normalem]{ulem}
\usepackage{booktabs}
\useunder{\uline}{\ul}{}
\usepackage{mathtools}
\usepackage{hyperref}
\usepackage{makecell}
\usepackage{tikz}
\usepackage{listings}
\usepackage{tablefootnote}
\usepackage{amssymb}
\usepackage{pifont}
\usepackage{dblfloatfix}
\usepackage{array}
\usepackage{float}
\usepackage{wasysym}
\usepackage{xcolor}
\usepackage{caption}

\begin{document}

\title{Enabling Beyond-Visual-Line-of-Sight Drones Operation over Open RAN 5G Networks with Slicing}


\author{Pau~Baguer,
		Esteban~Municio,
		Gines Garcia-Aviles, 
        and~Xavier~Costa-Pérez 
\IEEEcompsocitemizethanks{\IEEEcompsocthanksitem Pau~Baguer, Esteban Municio and Gines Garcia-Aviles are with i2CAT Foundation.
\IEEEcompsocthanksitem Xavier~Costa-Perez is with i2CAT Foundation, NEC Labs Europe and ICREA.}
}

\maketitle

\begin{abstract}
Among the foretold claims of the transition from 5G to 6G, Beyond-Visual-Line-of-Sight (BVLoS) drone operation has emerged as a prominent Internet-of-Robots enabler. However, safety concerns have been raised since BVLoS imposes strict requirements on performance and dependability on the technology, and requires robust regulatory frameworks. While current 5G technologies promise to meet the performance requirements in terms of throughput and latency, there is a lack of studies regarding how to achieve full reliability in practice. To address this challenge, the research community is actively working on open-source projects that allow for experimental validation in the field. 
Fortunately, new Open RAN (O-RAN) standards are paving the way for such approaches in an integrated, native manner.
In this work, we deploy a state-of-the-art 5G O-RAN open-source BVLoS operational system, report current limitations, and address them via advanced capabilities natively available in O-RAN: Slicing. Our proposed deployment minimizes trajectory errors due to 5G link congestion and keeps latency well below the 3GPP limits defined for BVLoS operation. Finally, we discuss on the challenges ahead and the opportunities that 5G O-RAN-enabled networks may bring to BVLoS drone operation.

\end{abstract}


\section{Introduction}
\label{sec:intro}

Unprecedented levels of network performance in terms of high data rates and low latency communication are being currently delivered in 5G networks. This has led to a significant increase in the use of 5G in Unmanned Aerial Vehicles (UAVs) or Drones. Despite this, current UAV use cases are mostly based on Visual Line-of-Sight (VLoS) operations, where direct visibility by the pilot is always ensured as a fail-over option in the case of an emergency. Beyond 5G (B5G) and 6G systems do promise Beyond VLoS (BVLoS) UAV operation use cases.
BVLoS flights are key to fully enabling U-space and Unmanned Aircraft System Traffic Management (UTM), two approaches that aim to safely integrate drones into the airspace of populated areas at low altitudes for new UAV applications including delivery services, infrastructure inspection, or air transport. However, there is still a long way to go before BVLoS use cases are accomplished in practice~\cite{geraci2022will}.

Although extensive regulatory frameworks and privacy-aware policies are being developed~\cite{schiller2023drone}, significant safety concerns have been set because of the strict network requirements and dependability levels imposed by BVLoS use cases. 
5G networks are in theory adequately capable of supporting such latency and throughput requirements, subject to sufficient coverage. 
However, there are only a few works such as \cite{muzaffar2020first} that provide practical experimentation results to back these figures. In addition to this, achieving five-nines reliability is still far on the horizon. Only next-generation B5G/6G cellular systems are expected to be able to reach it on a commercial, carrier-grade basis.

New generation cellular systems rely on new Open RAN (O-RAN) architectures, as the one being standardized by the O-RAN Alliance, to presumably enable such dependability levels thanks to a new flexible, disaggregated, and open approach that may truly materialize network softwarization in a multi-vendor RAN~\cite{garcia2021ran}. This means AI/ML approaches may be natively integrated into an architecture that offers open interfaces to deploy data-driven solutions for, among others, trajectory prediction, coverage optimization, traffic forecasting, or mobility management in LoS. 
 Also, such architecture opens up the RAN experimentation to third parties and open-source projects, unleashing a new level of innovation in the RAN space. 
However, while such architecture does address some of the aforementioned challenges, little is known about the real-world 5G performance to enable BVLoS, especially when leveraging on current open-source projects to assess their effectiveness in practical BVLoS scenarios.
 
In this work we evaluate an open-source O-RAN 5G-enabled drone operation deployment, highlighting the key O-RAN and 5G-NR 
BVLoS enablers. In our study, we point out the limitations that may hinder wide adoption (e.g., the impact of QoS degradation caused by network congestion in BVLoS drone control) and provide solutions that are natively available within current 5G O-RAN enabled architectures such as the E2 Service Model (E2SM) based network slicing. We also argue on the need to fulfill 3GPP Unmanned Aerial System (UAS) requirements for BVLoS as specified by~\cite{3gppUAS}, which means a latency communication lower than \textit{40 ms} for both telemetry and steering commands, and lower than \textit{140 ms} for First Person View (FPV) video for drone control.

\begin{figure*}
\centering
\includegraphics[width=1\textwidth]{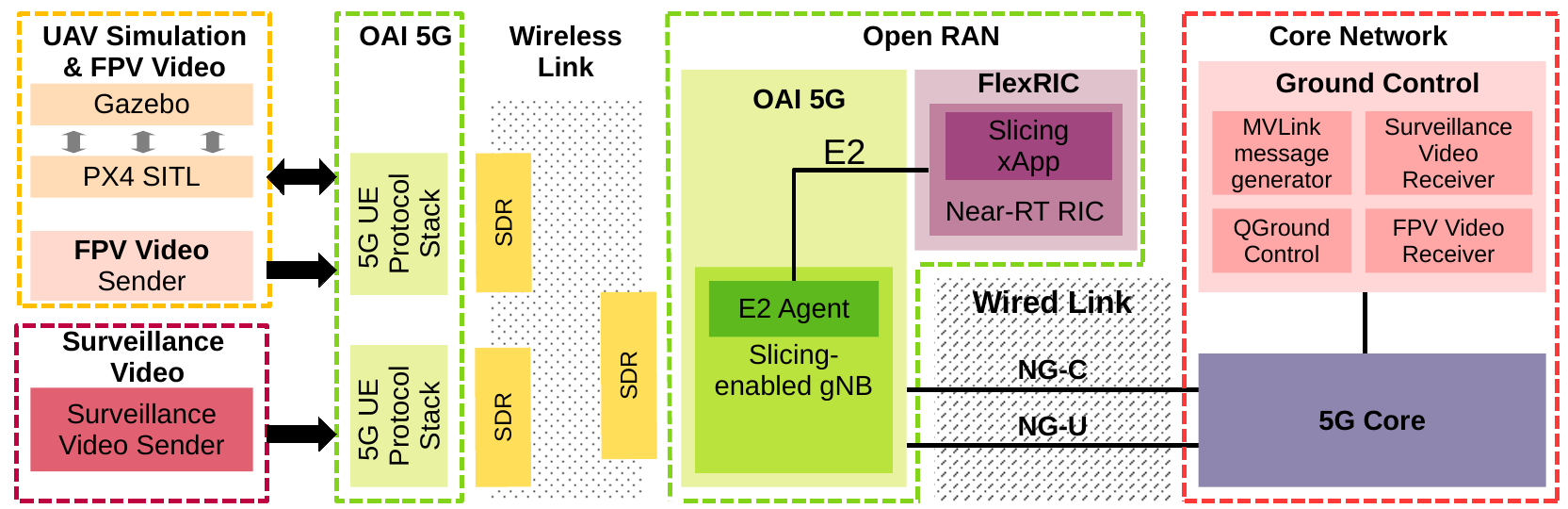}
\caption{System architecture depicting the components that form the deployment for testing an O-RAN enabled 5G network for BVLoS drone operations.}
\label{fig:deployment}
\vspace{-4mm}
\end{figure*}

Our experimental tests report that, unlike \textit{vanilla} Proportional Fair scheduling approaches, when network slicing is used, trajectory deviation attributable to the 5G link is drastically reduced. We also show that applying network slicing in state-of-the-art open-source 5G O-RAN enabled solutions can render overall latency values below the boundaries specified by 3GPP, although additional mechanisms may be needed to completely remove outliers and ensure full 3GPP compliance. Finally, we discuss the challenges that lay ahead and report on the lessons learned while deploying such an open-source 5G O-RAN-enabled solution for dependable BVLoS drone operation.   

Therefore, our main contribution is to integrate, evaluate, and characterize an open-source O-RAN-enabled 5G network, including the design and implementation of network slicing capabilities to demonstrate its feasibility for supporting BVLoS operation.

\section{Background}
\label{sec:background}

In this section, we first introduce the state-of-the-art in cellular-based drone control and discuss relevant works that paved the way towards BVLoS drone operations. Secondly, we introduce the main architectural enablers for Beyond 5G network architectures towards reliable BVLoS 
drones Internet of Robots use cases.

\subsection{Cellular-Based Drone Operations}
\label{subsec:oran_architectures}

Over the past decade, the use of VLoS-operated drones leveraging 4G cellular connectivity has witnessed remarkable progress across diverse industries, including aerial photography, maintenance surveillance operations, precision agriculture, and medical package delivery. The capacity to remotely control drones within the pilot's VLoS while enabling the transmission of data at high speeds has opened up new opportunities for live-feed data collection and real-time video monitoring in a cost-effective manner. 
However, although 4G may provide decent latency figures below 40 ms one-way, it still suffers from high jitter~\cite{donevski2021experimental}, throughput is usually limited to tens of Mbps, and full reliability remains an issue. More precisely,~\cite{baltaci2022analyzing} reports that 4G misses in providing enough reliability to maintain both low latency communication and high-quality video streaming simultaneously. This, together with the already known concerns on interference at altitude makes 4G unsuitable for BVLoS drone operations~\cite{ivancic2019flying}.

To overcome these limitations, 5G offers substantially higher data transfer rates, reduced latency, and improved reliability. Additionally, 5G 
promises seamless and real-time communication between cells, partially removing the high-latency spikes appearing in 4G handovers. 
Preliminary works on 5G BVLoS-operated drones such as~\cite{muzaffar2020first} confirm high data rates and low latency. However, authors also claim i) that 5G connectivity may not always be maintained, ii) can be affected by interference from other 5G base stations serving multiple airborne drones in the same frequency, and iii) may be highly dependent on the traffic present in the network.
Such traffic is expected to significantly increase in UTM/U-space systems due to a higher number of operating UAVs and, e.g., new edge computing approaches that may offload drone on-boarded tasks to the edge~\cite{hayat2021edge}. In such context of limited resources, end-to-end latency in 5G operated drones have been studied in~\cite{festag2021end} and~\cite{larsen2022xhaul}, where URLLC slices and Time Sensitive Networking (TSN) mechanisms are proposed for both the RAN and the fronthaul respectively. The former proposes a commercial network slicing implementation which 
relies on 5G NSA and the latter, 
although studies the latency of different
deployment options, it only provides non-experimental results.

Finally, the O-RAN architecture has been proposed to overcome some of the existing 5G limitations and to open up the BVLoS use cases. Through an O-RAN-enabled 5G network, drones may safely and seamlessly roam over different coverage areas, and high reliability may be ensured through natively implemented network slicing and 
AI/ML mechanisms from different vendors/developers using open interfaces that can be integrated by design in the RAN architecture. Some of the advantages of O-RAN in drone operation use cases are identified in~\cite{bertizzolo2021streaming}, where an O-RAN-enabled close-loop control system for drones is proposed. The system is able to exploit the O-RAN KPIs and control mechanisms to jointly optimize the location of the drone and the transmission directionality of the gNBs. Another example is \cite{pham2022ran}, where different learning methods are integrated into O-RAN to serve offloading tasks. Consequently, the O-RAN Working Group 1 (WG1) 
defines reference use cases for radio resource allocation based on flight path or the application scenario. Also, current O-RAN-enabled testbeds pay particular attention to controlled experimentation with aerial clients~\cite{mushi2023open}.

Unlike previous studies, in this work, we deploy a fully open-source O-RAN-enabled experimental 5G network aiming to fulfill the strict reliability requirements of a BVLoS drone operation.


\subsection{Beyond 5G Drone Architectural Enablers for BVLoS}
\label{subsec:oran_architectures}

B5G networks are expected to bring more advanced features and capabilities compared with current 5G architectures not only in terms of data rates and latency but also increased interoperability, flexibility, security, and privacy. Open RAN architectures are therefore valuable enablers for the development and deployment of future-generation networks given that their design principles build on top of openness, flexibility, interoperability, and intelligence. Intelligence consists of using new techniques based on AI/ML to efficiently optimize different RAN metrics and user QoE through multi-timescale control loops and effective data collection mechanisms. Additionally, 5G features such as network slicing are expected to be even more sophisticated to effectively enable the coexistence of a large variety of services on the same infrastructure. Hence, O-RAN architectures perfectly cope with the requirements for BVLoS drone operation by means of the following enablers:

\begin{itemize}
    \item[-] Closed-Loop Operation: Open RAN architectures include a key element responsible for making intelligent and dynamic decisions to optimize the network performance: the RAN Intelligent Controller (RIC). The O-RAN Alliance standardizes two different RICs operating at two different time scales, the Near-Real-Time RIC (near-RT RIC) operating between one and ten milliseconds and the non-Real-Time RIC (non-RT RIC), operating from tens to hundreds of milliseconds.
    \item[-] Native Network Slicing: Network slicing addresses isolation and resource allocation which is critical to address deterministic drone behavior while it is out of the vision field of the operator. O-RAN enables this natively through the E2 interface and the E2SM, which effectively facilitates the RAN control by allowing the near-RT RIC to execute policies directly on the gNBs within a near-RT control loop. 
\end{itemize}

\begin{figure*}
    \centering
    \begin{subfigure}[b]{0.485\textwidth}
        \includegraphics[width=\textwidth]{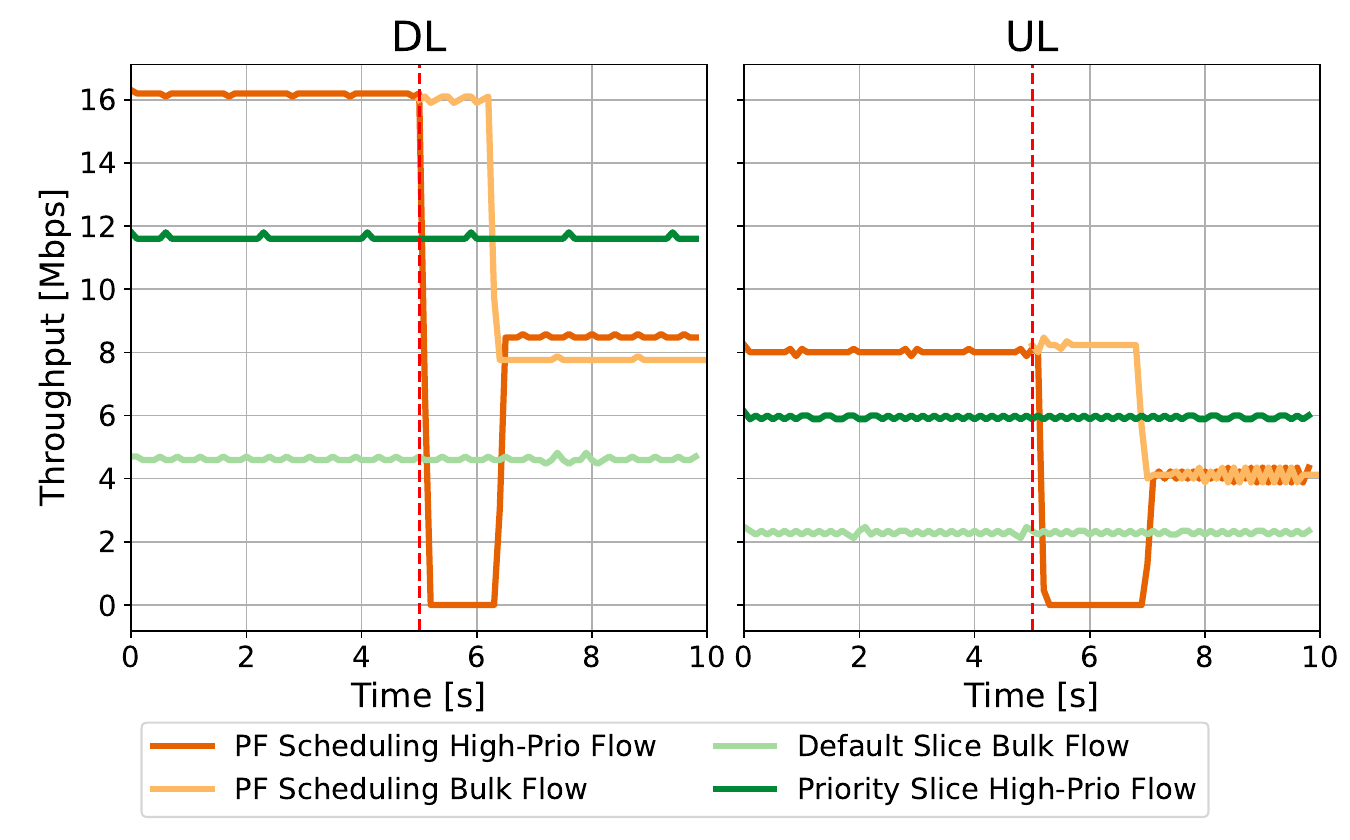}
        \caption{Throughput}
        \vspace{-2mm}
        \label{fig:gull}
    \end{subfigure}
    \begin{subfigure}[b]{0.505\textwidth}
        \includegraphics[width=\textwidth]{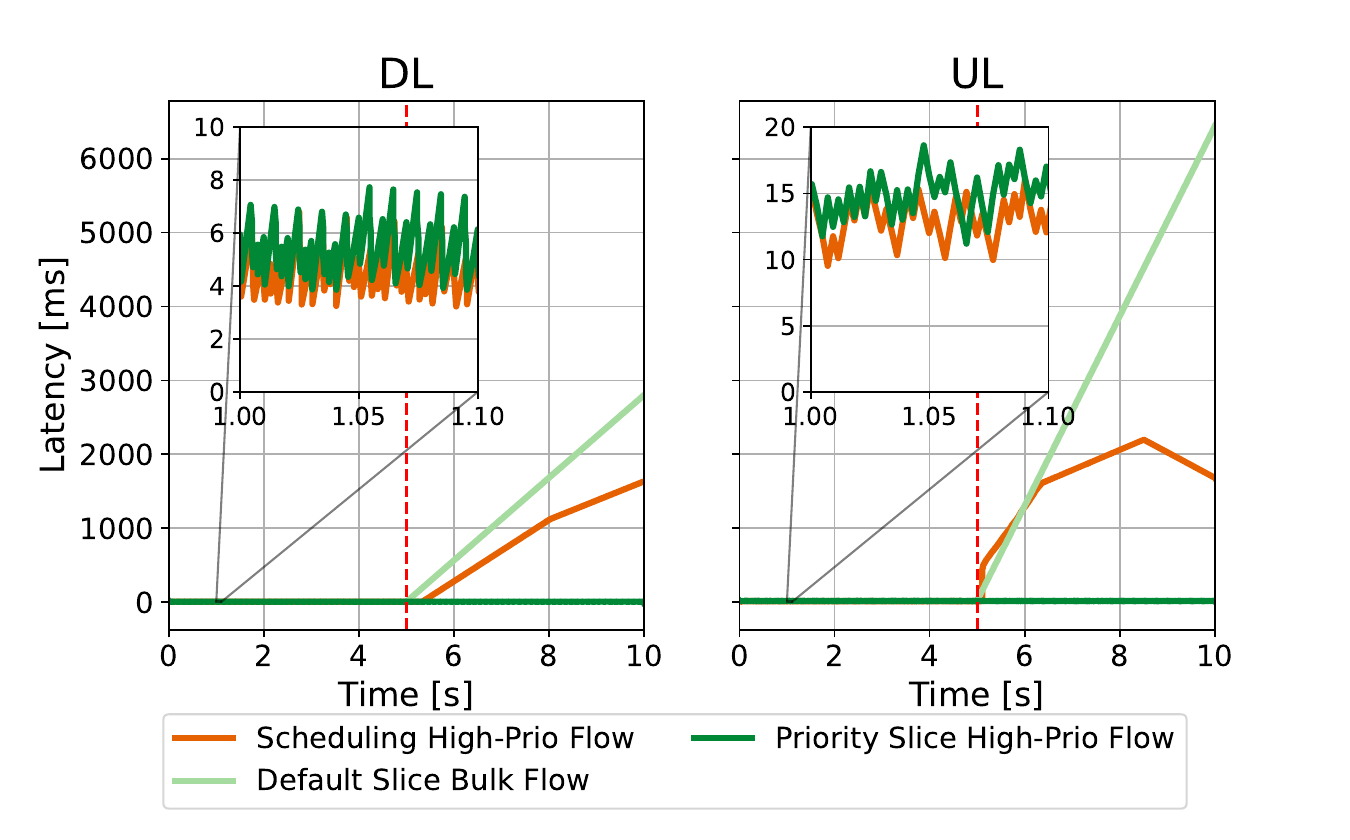}
        \caption{Latency}
        \vspace{-2mm}
        \label{fig:tiger}
    \end{subfigure}
    \caption{Slicing-enabled 5G link characterization}
    \label{fig:5g_slicing_characterization}
    \vspace{-6mm}
\end{figure*}

\section{Deploying BVLoS Drone operation through O-RAN enabled 5G networks.}
\label{sec:deployment_and_scenario}

\subsection{Experimental Deployment}
\label{subsec:deployment}

In order to experimentally evaluate the feasibility of a BVLoS drone operation solution based on an open-source O-RAN enabled 5G network, we have deployed a laboratory testbed based on  MOSAIC 5G FlexRIC, 
Open Air Interface (OAI) and PX4. Unlike other O-RAN near-RT RIC implementations such as the ones included in O-RAN Software Community (O-RAN SC) or the Open Networking Foundation’s (ONF) SD-RAN, FlexRIC 
supports bare-metal execution, allowing for fast execution times and low processing overhead. Additionally, it has a more mature xApp SDK with a wider set of available xApps. On the other hand, OAI is the 5G NR open-source reference implementation that has more advanced capabilities, including both TDD and FDD bands and up to 100 MHz of bandwidth. The 5G NR RAN is connected with a 5G core instance of the OAI-5G-CN project, implementing the different functional blocks present in a 5G Core Network for Standalone operation.
Finally, PX4 is an open-source autopilot flight stack based on \textit{MAVLink} widely used to control a number of aircraft, ground vehicles, and underwater vehicles. PX4 supports Software in the Loop (SITL) simulation, which allows the user to run the flight code in a computer-modeled vehicle as in a controlled environment.

Figure~\ref{fig:deployment} depicts the deployed architecture including 
the UAV (left), the 5G RAN (center), and the core network (right). The near-RT RIC is deployed to control the gNB via the E2 interface and an xApp is onboarded to provide slicing logic based on the information given by an external service (e.g., a non-RT RIC control loop or other E2 node KPIs). We use \textit{OAI}'s E2 Agent to access the E2 interface for direct communication with the near-RT RIC. Additionally, we have designed and implemented a slot-based network slicing procedure within OAI, exposing it as an E2 feature that can be controlled from the near-RT RIC. We configure the drone with two independent 5G TDD links assigned to different slices, one for drone remote control and another for user data. We use \textit{Ettus USRP B210} and \textit{B205} for the two 5G links of the drone, and an \textit{Ettus USRP N310} for the gNB operation.


\subsection{Testing Methodology}
\label{subsec:deployment}

To conduct a thorough assessment of the system described in Section~\ref{subsec:deployment}, we examine a practical scenario inspired by the \textit{Radio Resource Allocation for UAV Application Scenario} outlined within the O-RAN WG1.
It consists of a surveillance scenario where two pilots are controlling the UAV. Pilot 1 controls the position and navigation of the vehicle without VLoS. Consequently, Pilot 1 has a C2 connection with direct stick steering and a First-Person View (FPV) live feed to aid control. On the other hand, Pilot 2 is monitoring the real-time live video feed for surveillance purposes. In such a scenario, we can assess the BVLoS capabilities via the performance of the remote control commands issued by the pilot, the telemetry generated at the drone, the FPV video to assist \mbox{Pilot~1}, and the best-effort surveillance video delivered to Pilot 2 (acting as bulk application network traffic load external to the UAV operation flows).

Therefore, and as Figure~\ref{fig:deployment} shows, the scenario considers three types of flows: i) a prioritized FPV video stream for remote control in Uplink (UL), ii) a lower priority surveillance video stream in UL, and iii) a bidirectional critical C2 message flow consisting in prioritized manual joystick commands sent in the Downlink (DL) to remotely control the drone from the QGround Control (QGC) station, and telemetry data and ACKs in the UL.
Additionally, for some of the tests, we will also consider bulk traffic generated with \textit{iperf} that represents network load in the 5G link caused by other UAVs or computing tasks offloaded to the edge. 

The UAV tests are designed as the scenario of maximum networking dependency on UAV control. The PX4 autopilot is receiving MAVLink commands of type \textit{Manual Control} (69), which carry the controller buttons, i.e., longitudinal (x), lateral (y), elevation (z), and heading (r), pressed by the pilot. The position values have a range of ± 1000 and \textit{Manual Control} commands are sent at a frequency of 25 Hz. A square-like, counter-clockwise flight path has been chosen to easily identify correlations between poor network conditions and trajectory deviations. Take-off and landing phases are excluded from the tests.

Using this scenario, we will first characterize the performance of the 5G links both in UL/DL. Then, we will assess the impact of anomalies in both trajectory and manual control commands. Finally, we will analyze the performance of the FPV video while operating the drone. We perform the tests with both network slicing and the default OAI's \textit{Proportional Fair} (PF) scheduling mechanism to evidence the effect of network slicing on QoS compliance.

\section{Evaluation of BVLoS operation through 5G}
\label{sec:results}

In this section, we first evaluate the performance of the 5G NR link for both UL and DL. Later on, we assess the effects of the wireless link performance on the drone trajectory and telemetry data and finally, on the video flows.

\subsection{Wireless Link Characterization}
\label{subse:5g_characterization}

To characterize the 5G NR link, we perform the experiment setup as explained above, but keeping the drone static and replacing drone real traffic with synthetic \textit{iperf} UDP traffic. The experiment starts by sending high-priority traffic and, at t=5s, an additional UDP bulk flow (e.g., competing background traffic) is added with the purpose of testing the behavior of the link. We perform such tests with and without a network-slicing-enabled O-RAN architecture. When disabled, the behavior of the RAN in terms of resource management is set by the scheduling policy which in the case of OAI is PF, referred to as \textit{"PF Scheduling"} in the subsequent sections. In contrast, network-slicing-enabled O-RAN architectures, referred to as Slicing, allow for a more flexible and fine-grained management of the available resources with an inherent isolation between traffic flows. In this study, the E2SM network slicing configuration involves allocating two-thirds of the available resources to high-priority traffic (i.e., Priority Slice High-Prio Flow), while one-third is reserved for other traffic flows (i.e., Default Slice Bulk Flow).

\begin{figure*}
    \centering
    \begin{subfigure}[b]{0.49\textwidth}
        \includegraphics[width=\textwidth]{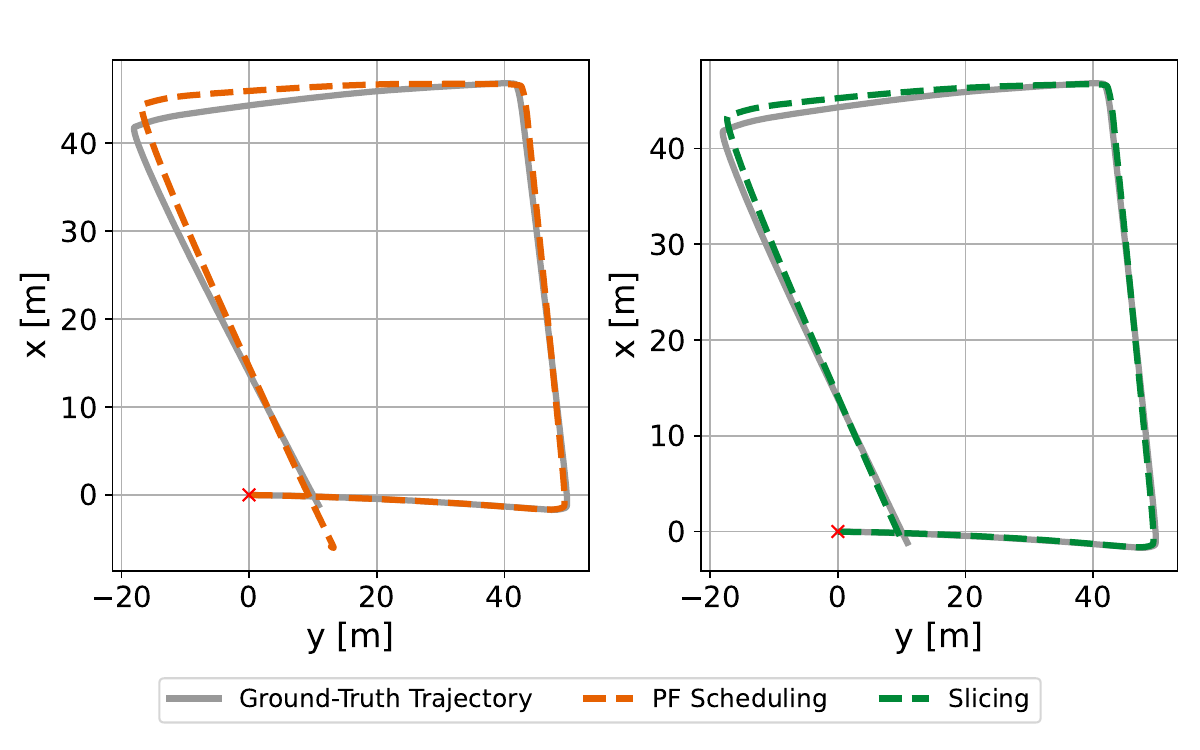}
        \caption{Drone trajectory}
        \label{fig:gull}
    \end{subfigure}
    \begin{subfigure}[b]{0.49\textwidth}
        \includegraphics[width=\textwidth]{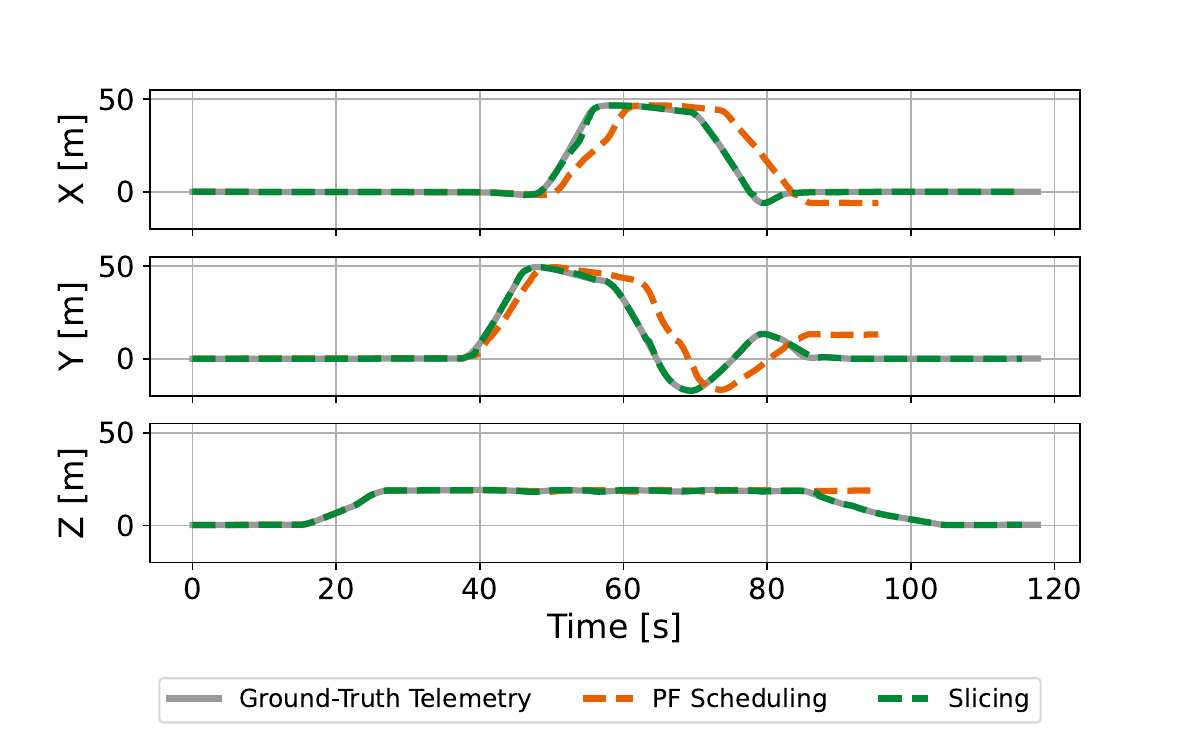}
        \caption{Telemetry traffic deviation}
        \label{fig:tiger}
    \end{subfigure}
    \caption{Effects of traffic performance on drone trajectory }
    \label{fig:trayectory_and_control_traffic}
    \vspace{-4mm}
\end{figure*}

Figure~\ref{fig:5g_slicing_characterization} shows the results of the experiment in terms of throughput and latency. For the case of DL throughput when network slicing is not enabled, PF
scheduling High-Prio Flow traffic uses all the link capacity (i.e., dark orange, $\sim$~16 Mbps on average) until at t=5s, the overall capacity is shared with the PF
scheduling Bulk Flow traffic (i.e., light orange, $\sim$~8 Mbps each) as we would have expected given the selected PF scheduling policy. Note that PF Scheduling uses a small transient window of about 1.5 seconds to compensate for the recorded overuse of the PF Scheduling High-Prio Flow, assigning temporarily all the resources to the PF Scheduling Bulk Flow. After 1.5 s, both flows share the channel equitably at 50\%. When network slicing is being used, both priority and background traffic start sending from t=0, and their corresponding throughput remains constant and proportionally isolated. Similar behavior can be observed for UL with the capacity adjusted to UL values. For the case of latency, a minor increase of latency is observed in the priority traffic when slicing is used because when slicing is implemented in a per-slot basis, there exist fewer TX opportunities compared to when no slicing is used (i.e., when using PF Scheduling). This can be observed in detail in Figure~\ref{fig:5g_slicing_characterization} b), where the Priority Slice High-Prio Flow latency (i.e., dark green) is slightly higher than in PF Scheduling High-Prio Flow (i.e., dark orange).
However, in t=5 when background traffic is added, the traffic scheduled by PF without slicing protection suffers a significant linear increase in latency. On the contrary, if the priority traffic is protected within a slice, the latency remains low and constant. For simplicity, the delay of the bulk flow when using the default scheduling policy (i.e., PF Scheduling Bulk Flow) is not included.

This first experiment evidences the need to protect priority traffic within 5G slices and reports latency values below 8 and 20 ms for DL and UL respectively. 

\subsection{Trajectory and Telemetry}
\label{subse:performance}

Once the link has been characterized, in this section we evaluate the effect of the network performance on the flight of the drone following the setup described in Section~\ref{sec:deployment_and_scenario}. The trajectory followed by the drone is depicted in Figure~\ref{fig:trayectory_and_control_traffic} a)
for the case when PF scheduling 
is applied (i.e., orange dashed line), and in Figure~\ref{fig:trayectory_and_control_traffic} b) when slicing is applied (i.e., green dashed line). Both trajectories are compared when the ground truth trajectory, i.e., the actual trajectory is exerted from the controller with negligible delay. A larger error is observed when using {\textit{PF Scheduling}}. This is because, as evidenced in Section~\ref{subse:5g_characterization}, the latency is higher when the priority traffic (i.e., the \textit{Manual Control} messages) is not protected from external background traffic. This results in a maximum error (as in Euclidean distance) of 2.4 m when RAN 
resources are managed by {\textit{PF Scheduling}} while it drastically decreases to a maximum of $1m$ when RAN {\textit{Slicing}} is being used.

Additionally, we also show in Figure~\ref{fig:trayectory_and_control_traffic} b) the effect of latency on the telemetry traffic, which reports the (x,y,z) location of the drone to the QGC. In BVLoS operation, such telemetry data is required to be as accurate as possible to allow the pilots to get a realistic control closed-loop.
When the default {\textit{PF Scheduling}} is selected, the location reported by the drone telemetry (i.e., orange dashed line) shifts with respect to the actual location (i.e., grey line), while when RAN slicing is enabled, the telemetry position (i.e., green dashed line) almost overlaps with the actual location. Again, this is because the slice protects the telemetry from other less prioritized external traffic, in this case, the video surveillance flow.

\subsection{BVLoS Operation Metrics}
\label{subsec:bvlos}

\begin{figure*}
    \centering
    \begin{subfigure}[b]{1\textwidth}
        \includegraphics[width=\textwidth]{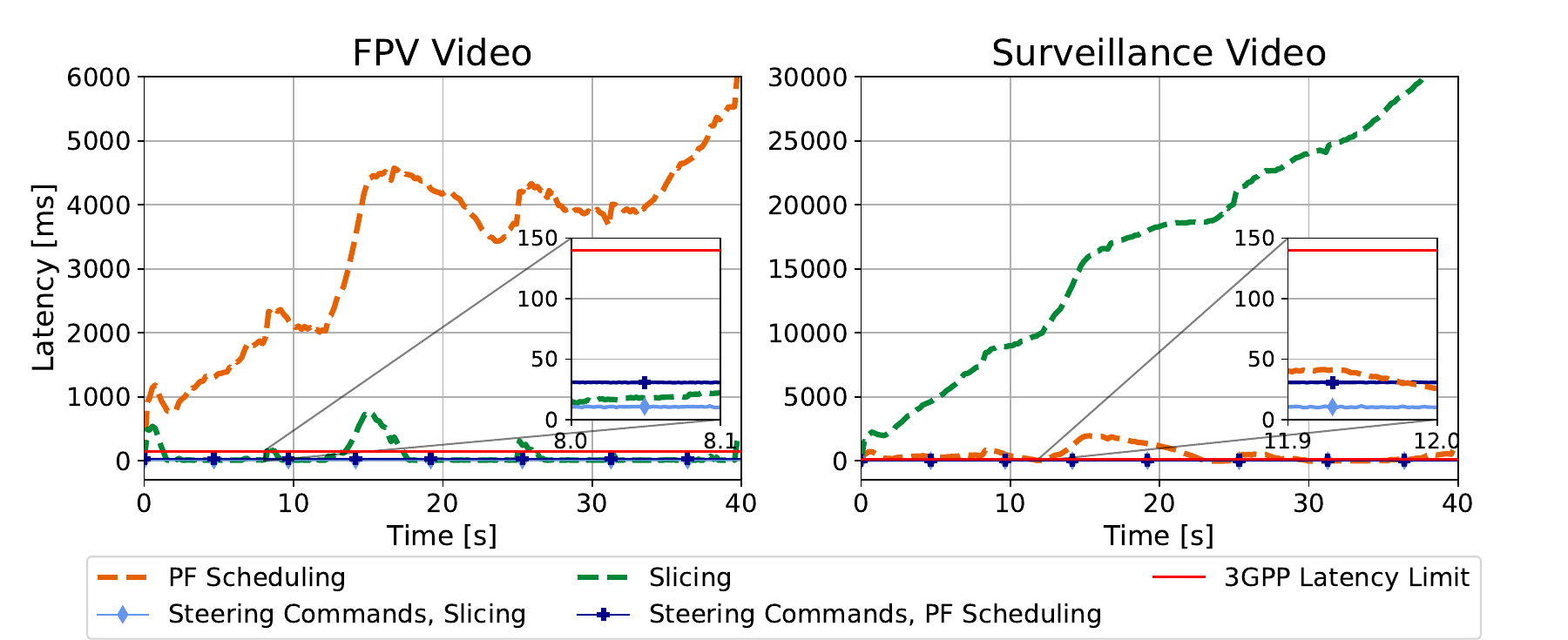}
        \label{fig:gull}
    \end{subfigure}
    \vspace{-4mm}
    \caption{Communication latency for FPV and Surveillance video in DL. 
    }
    \label{fig:latency-bounds}
    \vspace{-4mm}
\end{figure*}

Finally, we proceed with the evaluation of the FPV Video received by Pilot 1 and the Surveillance Video received by Pilot 2. Figure~\ref{fig:latency-bounds} depicts the latency of both FPV Video (left) and surveillance video (right) for both \textit{PF Scheduling} and \textit{Slicing} configurations. On the one hand, the FPV Video suffers an irregular increase in latency when network congestion appears if no network slicing technique is enabled. In contrast, network slicing brings FPV Video latency to the minimum and stabilizes its average value at 25 ms, well below the 140 ms required by 3GPP UAS 
due to the inherent isolation provided. This however does not always hold true, since bursty FPV behaviors may deplete the capacity of the slice and sporadically trigger the latency over the threshold (i.e., see the small latency peaks in the green dashed line in Figure 4 for FPV video). These effects can be smoothed and even removed by using a different video encoder or by applying online self-optimization mechanisms to adjust the resources assigned to different slices. For example, deep learning approaches such as~\cite{bega2019deepcog} report that it is possible to stay below a certain threshold with high probability with an over-provisioning of between 5\% and 20\%. On the other hand, the Surveillance Video latency increases when network slicing is enabled due to the limitations imposed to preserve the slice carrying FPV Video. Note that this would impede real-time surveillance use cases if the link capacity is low, as latency is heavily affected by the queuing delay and grows to the order of tenths of seconds. 
Finally, we have also included in Figure~\ref{fig:latency-bounds} the latency obtained for the steering commands in comparison with the latency of both FPV Video and the Surveillance Video.
We see that steering commands are not affected by any of the videos, since commands are sent in DL and videos in UL. Also, we observe that steering commands have a stable latency of about 10 ms (with slicing), and 30 ms (without slicing). This is because in both cases, steering commands are prioritized and protected, and when using slicing, there are more transmission opportunities which makes its latency even lower.

\section{Conclusions and Final Remarks}
\label{conclusions}

The transition from 5G to 6G networks will open up unprecedented possibilities for Beyond-Visual-Line-of-Sight (BVLoS) drone operations in the Internet of Robots use cases. While 5G and beyond technologies promise to offer high data rates and low latency, meeting the requirements to ensure full dependability in BVLoS scenarios remains a significant concern. However, with the advent of new Open RAN (O-RAN) standards, the research community is making quick strides towards enabling BVLoS over 5G and achieving full dependability in BVLoS drone operations.

In this work, we deploy and evaluate a state-of-the-art open-source 5G O-RAN-based solution for BVLoS drone operations, identifying the impact of network congestion on the QoS degradation 
and proposing closed-loop operation and network slicing to mitigate such effects. Our results show that current open-source 5G O-RAN-enabled solutions achieve overall communication latencies well below the 3GPP UAS requirements on real Software Defined Radio (SDR) hardware. Despite this, we observed video encoding asymmetries cause the communication latency to temporarily exceed such thresholds. On the other hand, we also observed on-lab deviations below 1 m in UAV simulation environments, setting a promising foundation for dependable BVLoS drone operations. 

Thus, there is still a long way to go before fully dependable BVLoS solutions may be achieved. In what follows, we report the lessons learned and  identify challenges to be addressed to unleash the full potential of BVLoS drone operations:

     $\cdot$ As we reported in Section~\ref{subsec:bvlos}, telemetry and FPV video may experience latency outbursts even when using network slicing if the 5G link capacity is limited. To remove such effects, bandwidth-stable or even adaptive video codecs may help. Additionally, AI/ML methods for dynamically adjusting the slice size to the actual demand can also reduce such latency outbursts and will result in a more efficient use of the 5G link capacity.
    
     $\cdot$ Slot-based RAN slicing, although lightweight, may lack enough granularity to provide prioritized traffic with a large number of TX opportunities. Resource Block (RB) based slicing can reduce latency in priority slices, although at the expense of a higher computing cost. Different BVLoS use cases may require different slicing strategies.
    
    $\cdot$ In order to further reduce control commands latency and improve the QoE in BVLoS operation, additional AI/ML techniques may be needed. Such techniques, which can be natively integrated into the O-RAN architecture as xApps, should target latency compensation, trajectory prediction, coverage optimization, traffic forecasting, or mobility management.

    Each of the previous remarks will result in different solutions to tighten the BVLoS control closed-loops which eventually will reduce the errors between virtual and actual trajectories. The research community may need to agree on standard metrics that objectively compare different BVLoS solutions. Among the most promising ones may be Dynamic Time Wrapping (DTW) or Fréchet distance to enable a fair comparison between virtual and real trajectories, or new BVLoS-specific QoE metrics that measure, from the pilot perspective, the relative difference between FPV video and received telemetry. 
    
\section{Acknowledgement}
  
This work has been supported by the European Commission through Grant No. 101017109 (DAEMON project), Grant No. 101097083 (BeGREEN project) and Grant No. 101139270 (ORIGAMI project), by the Spanish Ministry of Economic Affairs and Digital Transformation and the European Union – NextGeneration EU, in the framework of the Recovery Plan, Transformation and Resilience (PRTR) (Call UNICO I+D 5G 2021, ref. number TSI-063000-2021-3-Open6G), by CDTI and the European Union NextGenerationEU/PRTR
within the call “Ayudas Cervera para Centros Tecnológicos 2023” under Grant 6G-DIFERENTE (CER-20231018), and by the CERCA Programme from the Generalitat de Catalunya.


 \bibliographystyle{IEEEtran}
 \bibliography{refs}
\begin{IEEEbiographynophoto}{Pau Baguer} completed a double engineering degree in aerospace systems and networking from the Polytechnic University of Catalonia (Barcelona) in 2023. He joined the AI-driven Systems group of i2CAT in June 2022 and became a junior researcher in 2023. Pau's research interests are in Open RAN, AI-enhanced systems, and UAS.

\end{IEEEbiographynophoto}

\begin{IEEEbiographynophoto}{Esteban Municio}
received his Ph.D. degree from the University of Antwerp at imec (Belgium) in 2020. He then joined imec as postdoctoral researcher for two years. Since January 2022, he has been with i2CAT, where currently he is a research scientist in the AI-driven Systems group. His research interests are in the field of programmable open 5G networks, Time-Sensitive Networking and ultra-reliable Industrial IoT.
\end{IEEEbiographynophoto}

\begin{IEEEbiographynophoto}{Gines Garcia-Aviles}
received his Ph.D. degree in Telematics Engineering from the University Carlos III of Madrid at the IMDEA Networks Institute. Since January 2021, he has been with i2CAT, where currently he is a research scientist in the AI-driven Systems group. Gines’ research interests are Open RAN, RAN virtualization, and NFV technologies for future networks.
\end{IEEEbiographynophoto}

\begin{IEEEbiographynophoto}{Xavier Costa-Pérez}
is an ICREA research professor, scientific director at the i2CAT Research Center, and head of 5G Networks R\&D at NEC Laboratories Europe. He has served on the Organizing Committees of several conferences, published papers of high impact and holds tenths of granted patents. He received his Ph.D. degree in telecommunications from the Polytechnic University of Catalonia, Barcelona, and was the recipient of a national award for his Ph.D. thesis.
\end{IEEEbiographynophoto}

\end{document}